\documentclass[ prd, amsmath,amssymb,nofootinbib]{revtex4}

\usepackage{graphicx}
\usepackage{amsmath, amssymb, graphicx, epsf}
\usepackage{dcolumn}
\usepackage{bm}
\usepackage{hyperref}
\usepackage{epstopdf}
\newcommand{\md}{\mathrm{d}}
\newcommand{\mpl}{m_{\mathrm{p}}}
\newcommand{\nc}{\newcommand}
\nc{\ba}{\begin{eqnarray}}
\nc{\ea}{\end{eqnarray}}
\newcommand\be{\begin{equation}}
\nc{\K}{{\bf k }}

\begin{document}

\title{ No  Large Scale Curvature Perturbations during Waterfall of  Hybrid Inflation}

\author{Ali Akbar Abolhasani$^{1,2}$}
\email{abolhasani(AT)ipm.ir}
\author{Hassan Firouzjahi$^{2}$}
\email{firouz(AT)ipm.ir}
\affiliation{$^{1}$
Department of Physics, Sharif University of Technology, Tehran, Iran}
\affiliation{$^{2}$ School of Physics, Institute for Research in Fundamental Sciences (IPM),
P. O. Box 19395-5531,
Tehran, Iran}

\date{\today}

\begin{abstract}
In this paper the possibility of generating large scale curvature perturbations induced from the entropic perturbations during the waterfall phase transition of standard hybrid inflation model is studied. 
We show that 
whether or not appreciable amounts of large scale curvature perturbations  are produced 
during the waterfall phase transition depend crucially on the competition between the classical and the quantum mechanical back-reactions to terminate inflation.
If one considers only the classical evolution of the system we show  that the highly blue-tilted entropy perturbations induce highly blue-tilted large scale curvature perturbations during the waterfall phase transition which dominate over the original adiabatic curvature perturbations.  However, we show that the quantum back-reactions of the waterfall field  inhomogeneities produced during the phase transition 
dominate completely over the classical  back-reactions. The cumulative quantum
back-reactions of very small scales tachyonic modes terminate inflation very efficiently
and shut off the curvature perturbations evolution during the waterfall phase transition.
This indicates that the  standard hybrid inflation model is safe under large scale curvature perturbations during the waterfall phase transition. 
\end{abstract}

\maketitle
\section{Introduction}

Inflation proved to be very successful both theoretically \cite{Guth:1980zm}
and observationally \cite{Komatsu:2010fb} as a theory of early universe. The simplest models of inflation consist of a scalar field which is minimally coupled to gravity. A period of acceleration expansion is obtained if the potential is flat enough to allow for the 
inflaton field to slowly roll towards its minimum. With sufficient tunings in the parameters of the model, one can achieve  60 number of e-foldings or more to solve the horizon and the flatness problems of the standard  cosmology.

Hybrid inflation \cite{Linde:1993cn, Copeland:1994vg} is an interesting model of inflation 
containing two scalar fields, the inflaton field and the waterfall field. In Linde's original hybrid inflation, the energy density during inflation is dominated by the vacuum while the inflaton is slowly rolling. The waterfall field is very heavy compared to the Hubble expansion rate during inflation, $H$, and it quickly rolls to its instantaneous minimum. The potential has the property that once the inflaton field reaches a critical value, $\phi=\phi_c$, the waterfall field becomes tachyonic triggering an instability and inflation ends quickly thereafter and the systems settles down into its global minimum.

Usually it is assumed that the waterfall field does not play any role in curvature perturbations during inflation and during phase transition. In this picture one basically borrows the technics and the results of single field inflationary models. That is, the super-horizon curvature perturbations, once they leave the Hubble radius, are frozen and remain unchanged until they re-enter the Hubble radius at a later time, such as at the time of CMB decouplings. 

Here we would like to examine this picture more closely. We would like to see if hybrid inflation is safe under large scale curvature perturbations during the waterfall phase transition.
If one considers only the classical evolution of the system, we show that during the phase transition  the highly-blue tilted entropy perturbations can induce large blue-tilted curvature perturbations on super-horizon scales which can completely dominate over the original adiabatic curvature perturbations. However, we show that the quantum back-reactions of the waterfall field inhomogeneities produced  during the phase transition become important before the classical back-reactions become relevant. We demonstrate that  the cumulative quantum back-reactions of the short-wavelength inhomogeneities are so strong that they uplift the tachyonic instability of the entropy perturbations and the curvature perturbations freezes.

Ideas similar to this line of thought, studying the amplifications of large scale curvature  perturbations during preheating, were studied in \cite{Taruya:1997iv}-\cite{Kohri:2009ac} and more recently in  \cite{Levasseur:2010rk}. 

The paper is organized as follows. In section \ref{hybrid} we review the basics of hybrid inflation and obtain the background evolutions of the inflaton and the waterfall fields.
 In section \ref{entropy} the entropy perturbations and in section
\ref{curvature} their effects on curvature perturbations are studied. In Section \ref{back} the classical non-linear back-reactions as well as the quantum mechanical back-reactions 
are calculated and are compared to each other. Brief conclusions and discussions
are followed in section \ref{conclusions}.

While our work was finished the work by Lyth \cite{Lyth:2010ch} appeared which has overlaps with our results. See also \cite{Fonseca:2010nk} which appeared shortly after our work.

\section{Hybrid Inflation  }
\label{hybrid}

Here we study the basics of hybrid inflation \cite{Linde:1993cn, Copeland:1994vg} and the background field dynamics.

\subsection{The Potential} 
The potential in standard hybrid inflation has the form  
\begin{equation}
\label{pot}
V(\phi,\psi) = \dfrac{\lambda}{4} \left( \psi^2 - \dfrac{M^2}{\lambda}\right)^2 + \dfrac{1}{2} m^2 \phi^2 + \dfrac{1}{2}g^2 \phi^2 \psi^2 \, ,
\end{equation}
where $\phi$ is the inflaton field, $\psi$ is the waterfall field and $\lambda$ and $g$ are dimensionless couplings. The system has a global minimum given by $\phi=0$ and $\psi= M/\sqrt{\lambda}$. Inflation takes place for $\phi_c < \phi < \phi_i$
where $\phi_i$ is the initial value of the inflaton field and $\phi_c = M/g$ is the critical value of $\phi$ where the waterfall field becomes instantaneously massless. During inflation $\psi$ is very heavy and is stuck to its instantaneous minimum $\psi \simeq 0$.
For $\phi<\phi_c$ the waterfall becomes tachyonic triggering an instability in the system which ends inflation abruptly. Soon after phase transition, the systems settles down to its global minimum  and inflation is followed by the (p)reheating phase.

As in Linde's realization of hybrid inflation  \cite{Linde:1993cn}, we consider the limit where the inflation is dominated by the vacuum. For this condition to hold one requires that
\begin{equation}
\label{vac-condition}
M^2 \gg \dfrac{\lambda}{g^2} m^2 \, .
\end{equation}

To solve the flatness and the horizon problem, we assume that inflation proceeds at least for about 60 number of e-foldings.  In the vacuum dominated limit the number of e-foldings
is given by
\begin{equation}
\label{Ne}
N_e\simeq \dfrac{2 \pi~M^4}{\lambda \mpl ^2m^2} \ln \left( \dfrac{\phi_i}{\phi_c}\right) \, ,
\end{equation}
where $m_P^2 = 1/G$ with $G$ being the Newton's constant. We assume that $ \phi_i $ is few times $  \phi_c$ so one can basically neglect the logarithmic contribution above.

To get the correct amplitude of density perturbations, one has to satisfy the COBE normalization
for the curvature perturbations ${\cal P_R} \simeq 2 \times 10^{-9}$.
The power spectrum of curvature perturbations is
\ba
{\cal P_R} = \dfrac{128 \pi}{3 \mpl ^6} \dfrac{V^3}{V^2_{\phi}} 
 \sim \dfrac{g^2 }{\lambda^3} \dfrac{M^{10}}{\mpl ^6~m^4} \, ,
\ea
where the relevant quantities are calculated at the time of Hubble radius crossing ($k= a H$) at 60 e-folds before the end of inflation. In this picture, it is assumed that the curvature perturbations are 
frozen once the modes of interest leave the Hubble radius, as have been treated in conventional analysis of hybrid inflation so far. Our main goal in this paper is to examine the validity of this assumption more closely.

We are interested in the limit where the waterfall field rolls rapidly to its global minimum once 
the instability is triggered. For this to happen, the absolute value of the $\psi$ mass should be much bigger than the Hubble expansion rate during phase transition so
\begin{equation}
\label{waterfall-condition}
M^3 \ll \lambda m \mpl ^2 \, .
\end{equation}

\subsection{The Background Fields Dynamics}
\label{background}

Here we study the classical evolutions of background fields $\phi$ and $\psi$ during inflation and phase transition, see also \cite{GarciaBellido:1996qt, Copeland:2002ku, Randall:1995dj} where somewhat similar analysis were carried out too. In subsection \ref{back-reaction} we study the quantum back-reactions to the the system in the Hartree approximation. 

 The equations of motion for $\phi$ and $\psi$ are
\ba
\ddot{\phi} + 3H \dot{\phi} + (m^2+g^2 \psi^2) \phi =0 \\
\ddot{\psi} + 3H \dot{\psi} + (-M^2+g^2 \phi^2+\lambda\psi^2) \psi =0 \, .
\ea
With the assumption of vacuum dominated potential, the Hubble expansion rate is nearly constant during inflation, $H=H_0 \equiv \sqrt{2 \pi/3 \lambda} \,  M^2 / \mpl$. It is more convenient to use the number of e-foldings as the clock, $d N = H_0 \, d t$ and the background fields equations are now written as 
\ba
\label{phi-c}
\phi'' + 3 \phi' + \left(\alpha +g^2  \dfrac{\psi^2 }{H_0^2} \right) \phi =0 \\
\label{psi-c}
\psi'' + 3 \psi' + \left(-\beta +g^2 \dfrac{\phi^2}{H_0^2}+\lambda \dfrac{\psi^2} {H_0^2}\right) \psi =0 \, ,
\ea
where the dimensionless parameters $\alpha$ and $\beta$ are defined as
\ba
\alpha \equiv \dfrac{m^2}{H_0^2}=\dfrac{3 \lambda m^2 \mpl^2}{2 \pi M^4} \simeq 
\frac{3}{N_e}
\qquad , \qquad   \beta \equiv \dfrac{M^2}{H_0^2}=\dfrac{3 \lambda  \mpl^2}{2 \pi M^2}\, ,
\ea
and the prime denotes the differentiation with respect to the number of e-foldings.

Using the rapid waterfall condition Eq. (\ref{waterfall-condition}) combined with the 
expression for the total number of e-foldings Eq. (\ref{Ne}) one obtains
\ba
\label{vd-ne}
\beta \gg N_e \, .
\ea
Similarly, one can check that $\alpha \beta \gg 1$.

To simplify the notation, we take the critical point as the reference point and define
 $n \equiv N -N_c$. We use the convention  that at the start of inflation for $\phi=\phi_i$, $N=0$,  at the time of phase transition $N= N_c$ and at the end of inflation $N= N_e$. With this convention $n<0$ before phase transition whereas $n>0$ afterwards. 
As assumed, the $\psi$ field is much heavier than $H_0$ during inflation so it rapidly rolls down to $\psi \simeq 0$ and  one can simply solve {{Eq.}} \ref{phi-c}
\ba
\label{phif}
\phi(n) \simeq \phi_c \exp \left(-r ~n \right)  
\ea
with
\ba
r = \left( \dfrac{3}{2}- \sqrt{\dfrac{9}{4}-\alpha}\right) \simeq \frac{\alpha}{3} 
\simeq \frac{1}{N_e} \, .
\ea
Equivalently, one also has
\ba
\label{nc}
N _c \simeq \dfrac{1}{r} \ln \left( \dfrac{\phi_i}{\phi_c}\right) \, .
\ea

\begin{figure}[t]
\centerline{\includegraphics[scale=.3]{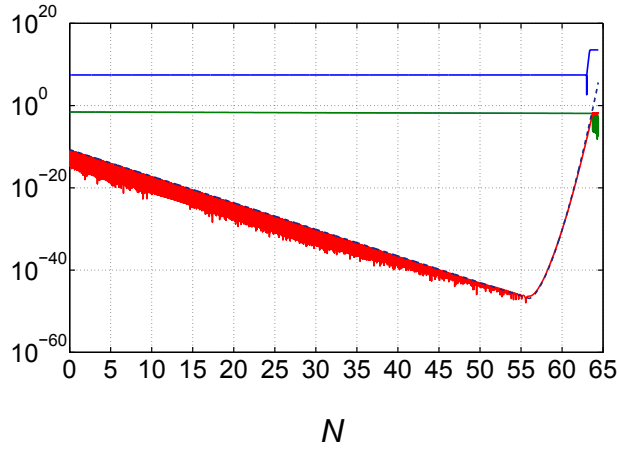}}
\caption{
The background classical dynamics for hybrid inflation potential { Eq.} (\ref{pot}) with
$M=.67 \times 10^{-7}$, $m=2.5 \times 10^{-10}$ and $\lambda = g^2 = 2.5 \times 10^{-11}$ in the units where $\mpl=1$. The upper solid blue curve shows the amplitude of curvature perturbations, the middle  solid green curve shows the background $\phi$ field evolution and the lower solid red curve shows the behavior of background $\psi$ field, all three curves are obtained from full numerical analysis. The dashed dark blue curve shows our analytical solution for $\psi$ field.  The vertical axis is logarithmic and the horizontal axis is the number of e-foldings.} \vspace{0.5 cm}
\label{bkd}
\end{figure}

Let us now turn to the dynamics of $\psi$ field. As it can be confirmed from our full numerical results we are in the limit where $\psi^2 / H_0^2 \ll \beta /\lambda$
so the equation for $\psi$ simplifies to
\ba
\label{psi-substituted}
\psi''+3\psi' +\beta \left( e^{-2r\,n} -1\right)\psi =0 \, .
\ea
We divide the solution into two regions. First, we solve this equation for the period before phase transition,  $\phi > \phi_c$. The solution of this equation during this period is 
\ba
\psi = e^{-3n/2} \left [ c_1 J_{\nu} \left( \dfrac{\sqrt{\beta}}{r} e^{-r~ n}\right)+c_2 Y_{\nu} \left( \dfrac{\sqrt{\beta}}{r} e^{-r~ n}\right) \right ] \, ,
\ea
where $J_\nu$ and $Y_\nu$ are the Bessel functions, $c_1$ and $c_2$ are constants of integrations  and $ \nu \equiv  \sqrt{\beta + 9/4}/r $.  As explained previously, we are in the vacuum dominated limit so $\beta \gg 1$ and
$\nu \simeq \sqrt\beta /r \simeq \sqrt\beta N_e \gg1$. Using the approximations for the Bessel functions with large arguments one finds
\ba
\label{psi-Nc1}
\psi \simeq e^{-(3-r)n /2 } ~ \left [ c'_1 \, \cos \left(\dfrac{\sqrt{\beta}}{r} e^{-r\,n} -\dfrac{1}{2} \nu \pi -\dfrac{1}{4} \pi \right) +c'_2 \, \sin \left( \dfrac{\sqrt{\beta}}{r} e^{-r\,n} -\dfrac{1}{2} \nu \pi -\dfrac{1}{4} \pi \right) \right ] \, .
\ea
This solution means that $\psi$ has oscillatory behavior with exponentially decaying frequency and exponentially decaying amplitude. This behavior can be seen in {\bf Fig. \ref{bkd}}.

Now, we solve the $\psi$ evolution for the period after phase transition till end of inflation, 
$\phi< \phi_c$.
Since inflation ends in few e-folds after $\phi$ reaches the critical point, it is more appropriate to use the small $r n$ approximation in { Eq.} (\ref{psi-substituted}) and 
\ba
\label{psi-eq-water}
\psi''+3\psi' -  (2\beta r\,n)\psi =0 \, ,
\ea
which has the following solution
\ba
\psi \simeq e^{-3n /2 } \left [ C _1 \mathrm{Ai}\left(\epsilon^{2/3}_{\psi}\,n + \dfrac{9}{4\epsilon^{4/3}_{\psi}} \right) + C _2 \mathrm{Bi}\left(\epsilon^{2/3}_{\psi}\,n + \dfrac{9}{4\epsilon^{4/3}_{\psi}} \right)  \right ] \, ,
\ea
in which 
\ba
\epsilon_{\psi}  \equiv \sqrt{2r\beta} \simeq \sqrt{\frac{2}{3} \alpha \beta} \, ,
\ea
and Ai(x) and Bi(x) are Airy functions of first and second kind respectively. For $\epsilon_{\psi} \gg 1$, which  is the case in our analysis to satisfy the water-fall condition, the second term in the argument of Airy function can be ignored. Since Ai(x) is a damping function for $x>0$ we just keep the term containing Bi(x). In a good approximation the solution can be read as
\ba
\label{orpsi}
\psi(n) \simeq  \psi(N_c) \, \mathrm{Bi} \left(\epsilon^{2/3}_{\psi} \,n\right)~ e^{-3n/2  } .
\ea
where $\psi(N_c) = \psi (n=0)$.
By using the asymptotic behavior of $\mathrm{Bi}(z)$ for large $z$, $z\rightarrow \infty$,
\ba
Bi(z) \propto \dfrac{e^{2/3 z ^{3/2}}}{\sqrt{\pi}\sqrt[4]{z}} \, ,
\ea
and keeping just the exponential dependence one finds that for $\phi < \phi_c$ 
\ba
\label{psi-2}
 \psi  \simeq \psi_i \exp \left(-\frac{3-r}{2} N_c \right)
 ~ \dfrac{1}{\epsilon^{1/6}_{\psi} n^{1/4}}~\exp \left( -\dfrac{3}{2}n +\dfrac{2\epsilon_{\psi}}{3} n^{3/2} \right)  ,
\ea
where to get the final result { Eq.}(\ref{phif}) have been used.


\begin{figure}[t]
\centerline{\includegraphics[scale=.3]{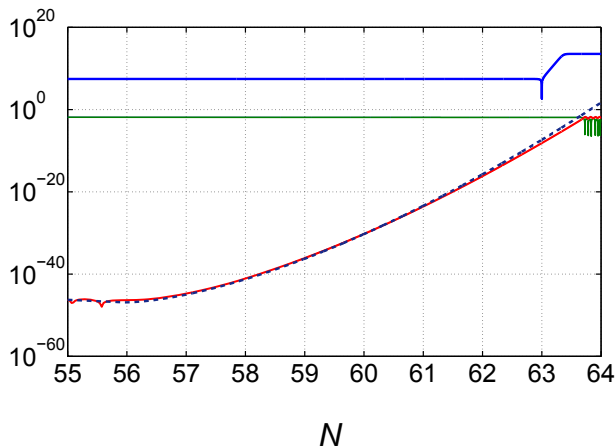}}
\caption{
Here the contents of  {\bf Fig. \ref{bkd} }  are shown in the last few e-foldings after the phase transition at $N_c \simeq 56$. Here we see that, considering only the classical fields dynamics, 
${\cal R}$ grows rapidly at the end of inflation until  the condition  (\ref{back-phi}) is met when ${\cal R}' \simeq 0$ at $N \simeq 63.5$. This is followed 
by  a short period of inflation when condition (\ref{back-psi})
is met and the non-linear corrections to $\psi$ dynamics settles it down to the global minimum ending inflation. }
\label{stf}
\vspace{0.5cm}
\end{figure}
 

\section{The entropy perturbations}
\label{entropy}

Here we look into entropy perturbation following \cite{Gordon00} closely.
In the field space of $(\phi, \psi)$ one can perform local fields rotations such that
\ba
\delta \sigma = \cos \theta \, \delta \phi + \sin \theta \, \delta \psi \quad , \quad
\delta s = -\sin \theta \, \delta \phi + \cos \theta \, \delta \psi \, ,
\ea
such that $\cos \theta = \dot \phi/\sqrt{\dot \phi^2 + \dot \psi^2}$ and 
$\sin \theta = \dot \psi /\sqrt{\dot \phi^2 + \dot \psi^2}$. In this picture, $\delta \sigma$ and $\delta s$ represent, respectively, the adiabatic and the entropic perturbations.  One can check that the evolution of the curvature perturbation for a mode with momentum $k$ is
\ba
\label{dotR}
\dot {\cal R} = \frac{H}{\dot H} \frac{k^2}{a^2} \Psi + \frac{2 H}{\dot \sigma } 
\dot \theta \delta s \, ,
\ea
where
\ba 
\label{dot-theta-eq}
\dot \theta = -V_s/\dot \sigma
\ea
with $V_s = \cos \theta \, V_\psi - \sin \theta \, V_\phi$.
 Eq. (\ref{dotR}) indicates that in the presence of large entropy perturbations or sharp turns in field space, the curvature perturbations can change on super-horizon scales.
 
The  equation of  entropy perturbations is \cite{Gordon00}
\ba
\label{deltas-eq}
\ddot{\delta s} + 3H \dot{\delta s} + \left( \dfrac{k^2}{a^2}+V_{ss}+3\dot{\theta}^2\right) \delta s = \dfrac{\dot{\theta}}{\dot{\sigma}}
\dfrac{k^2}{2\pi G a^2} \Psi \, ,
\ea
in which $V_{ss}=(\sin^2 \theta)^2 V_{\phi \phi}-(\sin 2\theta) V_{\phi \psi} +(\cos^2 \theta)^2 V_{\psi \psi}$.

\begin{figure}[t]
\centerline{\includegraphics[scale=.3]{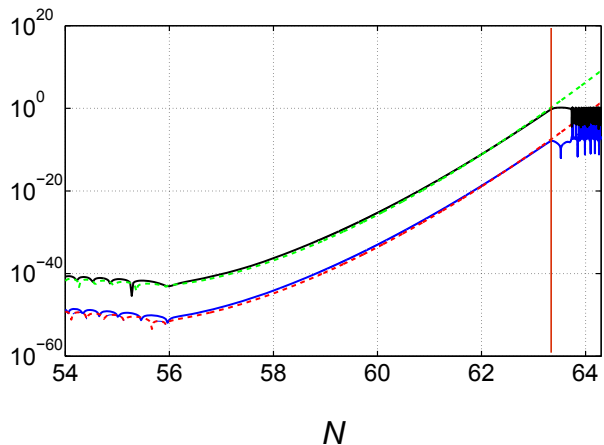}}
\caption{
Dynamics of angle in the $(\phi, \psi)$ field space with the same parameters as in 
{\bf Fig. \ref{bkd} }: The upper solid black curve and the lower solid blue curve, respectively, show the full numerical solutions of $| \ln \theta| $ and $ | \ln \theta'|$.  The upper dashed green curve and the lower dashed red curve, respectively, show the evolution of the corresponding quantities 
obtained from our analytical solutions, Eqs. (\ref{theta}) and (\ref{thetap}). Inflation ends when $\theta \sim 1$ indicated in the graph by the vertical line.
}
\label{bkt}
\end{figure}


We are interested in solving this equation for $\theta',  \theta \ll 1$. As we shall see later, these conditions are true until the end of inflation which can also be seen from {\bf Fig. \ref{bkt}}.
For the later references, from { Eq.} (\ref{phif}) and { Eq.}(\ref{psi-2}), one can easily find that
\ba
\label{theta}
\theta  \simeq \tan \theta =  \frac{\psi'}{\phi'} =
-\dfrac{1}{r} \left(\dfrac{\psi_i}{\phi_i} \right) \exp \left( -\dfrac{3-3r}{2} N_c \right)  \left(\epsilon^{5/6}_{\psi} n^{1/4} \right) ~\exp \left( \dfrac{2 \epsilon_{\psi}}{3} n^{3/2} - \dfrac{3-2r}{2}\,n \right) \, ,
\ea
which for $n>0$ or $\phi < \phi_c$ (after phase transition)  results in:
\ba \label{thetap}
\theta' \simeq \tan \theta' = -\dfrac{1}{r} \left(\dfrac{\psi_i}{\phi_i} \right) \exp \left( -\dfrac{3-3r}{2} N_c \right)  \left(\epsilon^{11/6}_{\psi} n^{3/4} \right) ~\exp \left( \dfrac{2 \epsilon_{\psi}}{3} n^{3/2} - \dfrac{3-2r}{2}\,n \right) \, .
 \ea

Neglecting $\dot \theta $ in Eq. (\ref{deltas-eq}), the equation governing the evolution of entropy fluctuations is the same as the equation of background $\psi$ field, Eq. (\ref{psi-substituted}). This indicates that, although the entropy perturbations are suppressed before the phase transition, but they becomes highly tachyonic during the phase transition. The subsequent tachyonic enhancement of the entropy perturbations overcome their initial suppression prior to phase transition.
This is the key effect to obtain large curvature perturbations induced from the entropy perturbations at the classical level. However, as we shall see in \ref{back-reaction},
one should also take into account the quantum back-reactions which can uplift the tachyonic instability of the background $\psi$ fields and modify the evolution significantly.

The differences between $\delta s$ and $\psi$ evolutions therefore are only in the initial conditions. For a given mode, we consider the evolution of the entropy fluctuation when it leaves the Hubble radius at the moment of Hubble radius crossing $N_*$,  so for $N>N_*$, similar to
Eq. (\ref{psi-2}), one obtains
\ba
\label{delpsi-n}
\delta s(n) \simeq \delta s_{\ast} ~  \left( \dfrac{\phi_i}{\phi_c}\right)^{1/2} ~ 
\exp\left( \dfrac{3-r}{2} N_{\ast} \right)  \exp(-\frac{3}{2}N_c)
 \exp \left( -\dfrac{3n}{2} +\dfrac{2\epsilon_{\psi}}{3} n^{3/2} \right) \, .
\ea
We note that the contribution $\exp\left(  2 \epsilon_\psi n^{3/2} /3  \right)$   represents the tachyonic enhancement of the entropy perturbations during the waterfall phase transition whereas the factors $\exp \left( -3 (N_c+ n)/2   \right)$ indicates the suppression of the entropy perturbations from the time of horizon crossing till the onset of phase transition. Both of these competitive behaviors can be seen in {\bf Fig. \ref {bkd}}.
 
To find the final amplitude of entropy perturbations, we need to find their amplitude at the time of Hubble radius crossing, $\delta s_*$.   To do this we note that the entropy perturbations, which are basically the waterfall field perturbations, are very heavy during inflation. As explained previously, this causes their suppression before the phase transition which should be taken into account.  During inflation
$\theta \ll 1$ so $\delta s = \delta \psi$. Rewriting the mode equation in conformal time 
$\eta = -1/aH$ and defining $v \equiv a(\eta) \delta s$ the equation of entropy perturbation is
\ba
\frac{\partial^2 v_k}{\partial \eta^2} + \left[ k^2 - \left(2- \dfrac{m^2_{\psi}(\psi)}{H^2} \right)\dfrac{1}{\eta^2} \right] v_k =0
\ea
in which 
\ba
\dfrac{m^2_{\psi}(\phi)}{H^2} \simeq \beta \left( \dfrac{\phi^2}{\phi^2_c} -1\right)  \, .
\ea
Since we are interested in momenta which exit the horizon during first few e-folds of inflation 
the above ratio is nearly constant and   
\ba
\dfrac{m^2_{\psi}(\phi)}{H^2} \simeq \beta \left( \dfrac{\phi_i^2}{\phi^2_c} -1\right) \equiv \tilde{\beta} \sim \beta \, .
\ea
By these considerations and noting  that $\beta \gg 1$ one has
\ba
\frac{\partial^2 v_k}{\partial \eta^2}  + \left[ k^2 + \dfrac{\tilde{\beta}}{\eta^2} \right] v_k =0 \, .
\ea
As the frequency of this equation changes adiabatically, one can use the WKB approximation  with the Bunch-Davis vacuum for initial times or $\eta \longrightarrow - \infty$ and obtains
\ba
\label{qem}
v_k \simeq \dfrac{1}{\sqrt[4]{4k^2+ 4\tilde{\beta}/\eta^2}}~ e^{\pm i \int\sqrt{k^2+ \tilde{\beta}/\eta^2} \mathrm{d} \eta } \, .
\ea
Using this solution the amplitude of the entropy perturbation at the time of Hubble radius  crossing, $k|\eta| =1$, is obtained to be
\ba
\delta\,s_{\ast} \simeq \dfrac{H}{\sqrt{2k^3}}\dfrac{1}{\tilde{\beta}^{1/4}}\Big \vert _{\ast} \, .
\ea
The extra factor $\tilde \beta^{-1/4}$ represents the suppression of the entropy perturbations before the phase transition.

Assuming that the scale factor at the start of inflation, $N_e$ e-folds before the end of inflation, is unity, $a(N_e) =1$, one also has
$N_{\ast} = \ln (a_{\ast}) = \ln \left( \dfrac{k}{H}\right)_{\ast}$. The final amplitude of entropy perturbations at the end of inflation therefore is
\ba
\label{delsf}
\delta s_f \simeq  ~\dfrac{1}{\tilde{\beta}^{1/4}} \dfrac{H}{\sqrt{2k^3}} \left( \dfrac{k}{H} \right)_{\ast}^{(3-r)/2}\left( \dfrac{\phi_i}{\phi_c}\right)^{1/2} e^{- 3/2 N_c}
~ \exp \left( -\frac{3}{2} n_f +\dfrac{2\epsilon_{\psi}}{3} n_f^{3/2} \right) \, ,
\ea
where $n_f \equiv N_e - N_c$ denotes the number of e-foldings from the start of phase transition till end of inflation. For a vacuum dominated potential with a quick phase transition, we have $n_f \sim 1$.

The final amplitude of the entropy perturbation ${\cal S} \equiv \frac{H}{\dot \phi} \delta s$ is
\ba
\label{Ps}
{\cal P_S} = {\cal A_S}  \left( \frac{k}{H_*}\right)^{(3-r)} \, ,
\ea
with
\ba
{\cal A_S}= \left(  \frac{H^2}{2 \pi \dot \phi} \right)^2 \dfrac{1}{\tilde{\beta}^{1/2}} \left( \dfrac{\phi_i}{\phi_c}\right) 
e^{- 3 N_c}
~ \exp \left( - 3 n_f +\dfrac{4\epsilon_{\psi}}{3} n_f^{3/2} \right) \, .
\ea
From Eq. (\ref{Ps} ) one observes that the entropy perturbations are highly blue-tilted
with $n_{\cal S} \simeq 4-r \simeq 4$. In next section we show that at the classical level 
these highly blue-tilted entropy perturbation induce  large blue-tilted spectrum on super-horizon 
curvature perturbations.


\section{Power Spectrum of Curvature Perturbations}
\label{curvature}

We now have all the materials to calculate the final power spectrum of curvature perturbations. For this purpose we need  to know the amplitude of adiabatic curvature perturbations at horizon crossing as the initial conditions and integrate the
evolutions of curvature perturbation from the time of horizon crossing till end of inflation.
The final amplitude of curvature perturbation, therefore,  is 
\ba
{\cal R}_f = {\cal R}_0 + \int_{0}^{n_f} {\cal R'} \mathrm{d}n \, ,
\ea
where ${\cal R}_0$ represents the adiabatic curvature perturbations in the absence of entropy perturbations. 

Starting with  Eq. (\ref{dotR}), the evolution of curvature perturbations for the super-horizon modes, induced by the entropy perturbations, can be written as 
\ba
\label{R-eq}
{\cal R}'= \dfrac{2 \, \theta'}{\sigma'}~ \delta \,s \, .
\ea
As one can see from above equation both $\theta'$ and $\delta s$ can source the curvature perturbations. We also note that $\theta'$ represents the acceleration of the field $\psi$, specially during the phase transition. As can be seen from the full numerical analysis, our classical background  is such that during inflation and phase transition, $\theta, \theta' \ll 1$. However, shortly after phase 
transition $\psi$ rises quickly from its value during inflation $\psi \simeq 0$ to its final value at the global minimum $\psi = M/\sqrt \lambda$ when inflation ends. 
The rapid rise of $\psi$ and $\psi'$ cause inflation to end when $\theta ,\theta'$ become large which can also be seen in {bf Fig. \ref{bkt}}. Equivalently, this can be interpreted as when the classical as well as quantum back-reactions from $g^2 \phi^2 \psi^2$ and 
$\lambda \psi^4$ interactions induce large masses for $\phi$ and $\psi$ such that they roll rapidly to the global minimum, ending inflation. Therefore, in the analysis below we work in the limit where $\theta, \theta' \ll 1$ and consider the end of inflation when $\theta = \theta_f \simeq 1$. 

To calculate the evolution of curvature perturbation from Eq. (\ref{R-eq}) we need to estimate
$\theta'$ and $\delta s$. The derivative of $\theta$ in field space is
\ba
\label{theta-eq}
\theta ' = \tan \theta ' = \dfrac{\psi''}{\phi'} - \dfrac{\psi ' \phi''}{\phi'^2} \, .
\ea
Since $r \simeq 1/N_e \ll 1$, the first term is  much larger than the second term by a factor of $~\epsilon_{\psi}/r$ and $\theta' \simeq \psi''/\phi'$.  To calculate $\delta s$ we observe that 
its equation has the same form as the background $\psi$ equation and therefore
\ba
\delta s (N) = \Omega _{s \psi} \psi(N) \, ,
\ea
in which
\ba
\label{Omega}
\Omega_{s \psi}\equiv \dfrac{\delta s(N_*)}{\psi(N_*)}=
\dfrac{\delta s_{\ast}}{\psi_i}~\left( \dfrac{k}{H}\right)_{\ast}^{\frac{3-r}{2}}
~ e^{i \delta \phi} \, .
\ea
The term $e^{i \delta \phi}$ represents the phase difference between the oscillations of $\psi$ and the entropy perturbations. This phase difference vanishes after time averaging when we find the final curvature power spectrum. 

Combining the above expressions for $\theta'$ and $\delta s$, the final curvature perturbation is integrated to
\ba
\label{RNe}
{\cal R} (N_e)= {\cal R}_0  -2 \int_{0}^{n_f} \dfrac{\Omega _{s \psi}}{\phi'^2} \psi''\psi ~\mathrm{d}n \, .
\ea
As the function $\mathrm{Bi}$ grows more rapidly than the linear exponential, from { Eq.} \ref{orpsi} one obtains
\ba
\psi''(n) \simeq  \psi(N_c)~e^{-3/2 n }~ \mathrm{Bi}'' \left(\epsilon^{2/3}_{\psi} \,n\right)  \, .
\ea
Using { Eq.} (\ref{sdaf}) and noting that  $\phi' \simeq  -r \phi_c \exp(-r~n)$ from 
{ Eq.} (\ref{phif}), the final curvature perturbations is obtained to be

\ba
\label{Rn}
{\cal R} = {\cal R}_0 - 2 \psi(N_c)^2~ \Omega _{s \psi} \dfrac{\epsilon_{\psi}^2}{r^2 \phi_c^2}\int_{0}^{n_f} n~e^{-(3-2r)\,n }~ \mathrm{Bi}^2\left(\epsilon^{2/3}_{\psi} \,n\right) ~\mathrm{d}n \, .
\ea

The amplitude of quantum fluctuations of adiabatic perturbations at the moment of horizon crossing is 
\ba
Q_{\sigma \ast} = \dfrac{H}{\sqrt{2\,k^3}} \Big \vert _{\ast} \, ,
\ea
where $Q_\sigma$ represents the Sasaki-Mukhanov variables for the adiabatic 
perturbations  \cite{Gordon00}. 
Using the form of $\Omega _{s \psi}$ given in Eq. (\ref{Omega}), the curvature perturbation calculated from Eq. (\ref{Rn}) is 
\ba
\label{Rnf}
{\cal R}(n_f) = {\cal R}_0 \left (1- {\cal C} \left(\dfrac{k}{H}\right)_{\ast}^{\frac{3-3r}{2}} {\cal I}(n_f) \right)   \, ,
\ea
where ${\cal R}_0$ is the initial value of adiabatic curvature perturbations at the moment of horizon crossing  
\ba
{\cal R}_0 = \dfrac{H}{\dot{\phi}}\dfrac{H}{\sqrt{2 k^3}} \Big \vert _{\ast} \, ,
\ea
and 
\ba
{\cal C} = \dfrac{2 \epsilon_{\psi}^2}{r\,\tilde{\beta}^{1/4}} ~ \dfrac{\psi_i}{\phi_i} e^{(-3+3r) N_c} \, .
\ea
Furthermore, the integral  ${\cal I}(n_f)$ has the following form
\ba
\label{In1}
{\cal I}(n_f) = \int_{0}^{n_f} n'~e^{-(3-2r)\,n' }~ \mathrm{Bi}^2\left(\epsilon^{2/3}_{\psi} \,n'\right) ~\mathrm{d}n' \, .
\ea

Eq. (\ref{Rnf}) has some interesting features. If the second term in the big bracket in Eq. (\ref{Rnf}) is larger than unity, then the induced curvature perturbations from the entropy perturbations dominate over the adiabatic curvature perturbations. Furthermore,  the dominant momentum dependence in the big bracket 
comes from the initial amplitudes of entropy perturbations and the integral  ${\cal I}(n_f)$ 
is nearly constant for all  momenta. This implies that the induced curvature perturbations from the entropy perturbations are highly blue-tilted. 

Using the integral approximation given by Eq.   ( \ref{mixint}) one can calculate the integral approximately and 
\ba
\label{In2}
{\cal I}(n_f) \simeq   \dfrac{\sqrt{n_f}}{2 \epsilon_{\psi}}~e^{-(3-2r)\,n_f }~ \mathrm{Bi}^2\left(\epsilon^{2/3}_{\psi} \,n\right)  \, .
\ea
Plugging this into Eq. (\ref{Rnf}), and using the asymptotic behavior of Airy function of second kind,  the curvature perturbations at the end of inflation is calculated to be
\ba
\label{Rnf2}
{\cal R}(n_f) = {\cal R}_0 \left [1- \left(\dfrac{k}{H}\right)_{\ast}^{\frac{3-3r}{2}}  \left(\dfrac{\epsilon_{\psi}^{2/3}}{\pi \,r\,\tilde{\beta}^{1/4}}\right)  \left(\dfrac{\psi_i}{\phi_i}\right) e^{(-3+3r) N_c} \exp \left( 4/3  \epsilon_{\psi} n_f^{3/2} -(3-2r)n_f  \right) \right]
\ea
To get the final curvature power spectrum we need to know $n_f$, the time of end of inflation. To determine this we proceed as follows.
The exponential growth of the $\psi$ modes  (background as well as quantum fluctuations) can violate our  background solutions for  $\phi$ and $\psi $. The exponential growth of the background $\psi $ field have two important effects. First, through the interaction term $g^2 \phi^2 \psi^2/2$, it can induce large mass for $\phi$ field which speeds up its rolling toward the global minimum and violate the slow-roll conditions. The second  important effect is that the back-reaction from $\lambda \psi^4 /4$ term induces large positive mass for $\psi$ which uplifts the tachyonic mass of $\psi $ field leading to the deceleration of this field. These two effects jointly terminate both inflation and the growth of the super-horizon curvature perturbations. To see the latter effect we start from { Eqs.} (\ref{theta-eq})  and  (\ref{R-eq}) where
\ba
{\cal R}' \sim \dfrac{\psi'' \psi}{\phi'^2} \, .
\ea
We observe that the deceleration of the $\psi$ field as well as the fast-rolling of the $\phi$ field jointly cause the termination of the super-horizon curvature perturbations as explained above. 


\section{Classical vs. Quantum Mechanical Back-reactions}
\label{back}

In this section we study the  back-reactions of classical (zero momentum) mode of $\psi$ field as well as the quantum back-reactions of $\psi$ inhomogeneities produced during phase transition on the dynamics of the system.  We examine which of the above two mechanisms dominate sooner to terminate inflation. 

We will demonstrate that the quantum mechanical back-reactions dominate completely over the classical back-reactions and hence the end-point of inflation is determined by quantum back-reactions. However, as we shall see later, the variance of the quantum fluctuations (on all scales) after phase transition has the same time dependence ($n$-dependence) as the classical trajectory and the difference is just in a proportionality factor.
In order to demonstrate why the classical back-reactions are secondary we start the analysis with the classical back-reactions.
\subsection{Classical Back-reactions}
\label{classical}

Due to smallness of background $\psi$ field before phase transition our analysis in \ref{background} concentrated only on the linear level. Now we add the back-reactions of 
the non-linear terms $g^2 \phi^2 \psi^2/2$ and $\lambda \psi^4/4$  on the dynamics of the system at the classical level.

The correction from the 
interaction term $g^2 \phi^2 \psi^2/2$ becomes important in $\phi$ evolution, { Eq.}(\ref{phi-c}), when
\ba
\label{back-phi}
\alpha \simeq g^2 \dfrac{\psi^2(n)}{H_0^2} \, .
\ea 
On the other hand, the non-linear self-interaction term in background $\psi$ evolution, {\ Eq}. (\ref{psi-c}), becomes important when 
\ba
\label{back-psi}
\beta \,r \simeq \lambda \dfrac{\psi^2(n)}{H_0^2} \, .
\ea
However, $ \lambda \alpha / g^2 \beta r \simeq \lambda/ \beta g^2 \ll 1 $ as a consequence of the vacuum domination condition { Eq.}(\ref{vd-ne}). This indicates that the back-reaction of the $\psi$ field on the inflaton dynamics becomes important sooner before its self-interaction corrections  affect its own dynamics. Putting it another way, during hybrid inflation 
$\epsilon \ll \eta$ where $\epsilon $ and $\eta$ are the conventional slow-roll parameters. Due
to back-reactions of $\psi$ on inflaton mass the condition $|\eta| \simeq 1$ is met sooner before $\epsilon$ find the chance to become order of unity due to its initial smallness during inflation.
We have numerically checked that when the condition Eq. (\ref{back-phi}) is satisfied then
$R' \simeq 0$. This followed by a very short period of inflation when condition Eq.  (\ref{back-psi}) is met and inflation ends.

Combining { Eq.}(\ref{back-phi}) and { Eq.}(\ref{psi-2}) one finds the end-point of inflation to be 
\ba
\label{c-back}
\exp \left( \frac{4}{3} \, \epsilon_{\psi} \,n_f^{3/2} -3 n_f  \right) \simeq \epsilon^{1/3}_{\psi} \left( \dfrac{\alpha}{\beta}\right)  \left(\dfrac{\phi_i }{\psi_i} \right)^2 e^{3 N_c}\, .
\ea
Using this in Eq. (\ref{Rnf2}), the amplitude of curvature perturbation at the end of inflation is
\ba
\label{Rnf3}
{\cal R}_f \simeq{\cal R}_0\left [ 1- \left( \dfrac{k}{H}\right)_{\ast}^{(3-3r)/2} \left(\dfrac{\epsilon_{\psi} \alpha}{ \pi \,r \beta \tilde{\beta}^{1/4}} \right) \left(\dfrac{\phi_i}{\psi_i} \right)  \right] \, .
\ea
Consequently, the final  power spectrum of curvature perturbations, ${\cal P_ R}  \delta^3({\bf k} - {\bf k'})= \frac{k^3}{2 \pi^2}  \langle {\cal R}({\bf k}) {\cal R}({\bf k'})  \rangle$,  
is calculate to be
\ba
\label{final-power}
{\cal P_ R} = {\cal P_R}_0\left [ 1+ \left( \dfrac{k}{H}\right)_{\ast}^{(3-3r)} \left(\dfrac{ \epsilon_\psi^2 \alpha^2}{2 \pi^2 r^2 \beta^2 \tilde{\beta}^{1/2}} \right)  \left(\dfrac{\phi_i}{\psi_i} \right)^2 \right] \, ,
\ea
where ${\cal P_R}_0 = k^3\,{\cal R}^2_0/2\pi^2$ is the power spectrum of the adiabatic curvature perturbations. 
As explained previously, during inflation $\psi$ is very heavy so $\psi$ rolls to its instantaneous minimum very quickly and $\psi_i \simeq 0$. Below we demonstrate that $\phi_i/\psi_i \gg 1$ such that the curvature perturbations induced from the entropy perturbations dominate completely over the adiabatic curvature perturbations. To see this note that 
for the $\psi$ field to be nearly zero such that its fluctuations  do not contribute to the curvature perturbation around $N_e$ e-foldings before the end of inflation, that is $\dfrac{{\cal R}'}{{\cal R}} \ll 1$ at the start of inflation,
one requires that 
\ba
\dfrac{{\cal R}'}{{\cal R}} \simeq \theta' \dfrac{\delta s_{\ast}}{\delta \sigma_{\ast}} \ll 1, \qquad \rightarrow \qquad \dfrac{\psi''}{\phi'} \dfrac{\delta s_{\ast}}{\delta \sigma_{\ast}} \ll 1 \, .
\ea
Using { Eq.}(\ref{psi-substituted}) and { Eq.}(\ref{phi-c}) one can easily find that 
\ba
\theta_i' \simeq\dfrac{\beta \, \psi_i}{r \, \phi_i}
\ea
This in turn results in 
\ba
\label{psi-phi}
\dfrac{\psi_i}{\phi_i} \ll \dfrac{r \tilde{\beta}^{1/4}}{\beta} \, .
\ea
Now compare the second term in the big bracket in Eq. (\ref{final-power}) to the initial curvature perturbations 
\ba
\frac{\Delta {\cal P_ R} }{{\cal P_R}_0} \equiv \left( \dfrac{k}{H}\right)_{\ast}^{(3-3r)} \left(\dfrac{\epsilon_\psi^2 \alpha^2}{2 \pi^2 r^2 \beta^2 \tilde{\beta}^{1/2}} \right) \left(\dfrac{\phi_i}{\psi_i} \right)^2 \, .
\ea
Noting that $(k/H)_{\ast} = e^{N_{\ast}}$ with $ N_* \gtrsim 1$ for super-horizon modes, and using the inequality Eq. (\ref{psi-phi}) and $r \simeq \alpha /3$ one obtains 
\ba
\label{DeltaR}
\frac{\Delta {\cal P_ R} }{{\cal P_R}_0} \gg \left( \frac{9\,  e^{3 N_{\ast}} }{\pi^2} \frac{\beta}{\tilde \beta}\right)\,  N_e .
\ea
Noting that $N_* \gtrsim1$ and $\tilde \beta \sim \beta$, one concludes that 
$\frac{\Delta {\cal P_ R} }{{\cal P_R}_0} \gg  N_e $. This indicates that, in the limit where only the  classical back-reactions  are considered to determine the endpoint of inflation, 
the curvature perturbations  induced from the entropy perturbations dominate completely over the original adiabatic curvature perturbations. We shall see in next subsection that this conclusion is not stable against quantum back-reactions.

\begin{figure}[t!]
\centerline{\includegraphics[scale=.3]{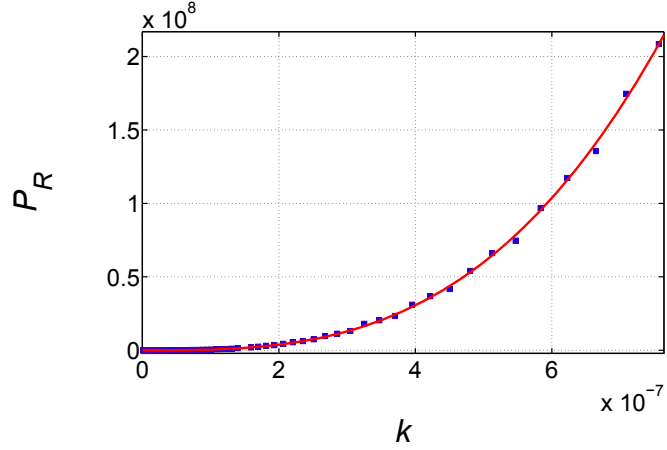}}
\caption{\label{spectral}
Here the final power spectrum of curvature perturbations, at the classical level,
as a function of comoving momentum for the 
parameters used in {\bf Fig. \ref{bkd}}.  are shown. The momenta exited approximately in first 6 e-folds. Blue squares shows the full numerical results whereas the solid red curve shows the best fit. The best power law fit is acquired for ${\cal P_R } = {\cal A} k^{n_s-1}$ with $n_s\simeq 4.01  (3.98-4.04 ) $ at $95   \%  \, \mathrm{CL}$. } 
\end{figure}

With  $\frac{\Delta {\cal P_ R} }{{\cal P_R}_0} \gg 1$, the spectral index of curvature perturbations, $n_{\cal R}-1 \equiv d \ln {\cal P_ R}/d \ln k$,  is
\ba
n_{\cal R} -1  = n^0_{\cal R} -1 + 3-3r  \, ,
\ea
in which $n^0_{\cal R}$ is the spectral index calculated from the adiabatic curvature perturbations using ${\cal P_ R}^0$, $n^0_{\cal R}-1 = 2 \eta -6 \epsilon \simeq 2 \eta = 2 \alpha/3$, where $\epsilon$ and $\eta$ are the standard slow-roll parameters. Using the approximation $r \simeq \alpha /3 \simeq 1/N_e$ one has
\ba
\label{nR}
n_{\cal R} -1  =  3 - \dfrac{1}{N_e}  \, .
\ea
This indicates that the curvature perturbations receive a large blue-tilted spectrum from the entropy perturbations and $n_{\cal R} \simeq 4- 1/N_e$.

Finally, to calculate the time of end of inflation, $n_f$, from Eq. (\ref{c-back}) we obtain
\ba
n_f \simeq \left[\dfrac{9}{4\epsilon_{\psi}}N_c + \dfrac{3}{4\epsilon_{\psi}}\,\ln \left(\epsilon^{1/3}_{\psi} \left( \dfrac{\alpha}{\beta}\right)  \left(\dfrac{\phi_i }{\psi_i} \right)^2 \right) \right]^{2/3}
\ea
For $(\phi_i/\psi_i) \sim 10^3 \beta/r $, the second term in the big bracket above can be ignored and  one approximately has
\ba
n_f \simeq \left(\dfrac{9}{4\epsilon_{\psi}}N_c \right)^{2/3} \, .
\ea

It is also interesting to calculate the angle in phase space at the end of inflation, $\theta_f$.
Using Eqs. (\ref{theta}) and (\ref{c-back}) we obtain
\ba
\theta_f \simeq 3 e^{-\frac{3}{2} r N_c}  \sim 1\, .
\ea
This verify our previous claim that inflation ends classically when $\theta_f \sim1$. This can also be seen from {\bf Fig} \ref{bkt}.

As an example consider the vacuum dominated hybrid inflation with parameters 
 $M=0.67 \times 10^{-7} \mpl$, $m=2.5 \times 10^{-10}\mpl$ and $ \lambda = g^2 = 2.5 \times 10^{-11}$.  From  Eq. (\ref{nR}) the spectral index  is 
$n_{\cal R} \simeq 3.99 $ which is in good agreement with $n_{\cal R}=4.01$ obtained from the full numerical analysis.  Also from {\bf Fig.} \ref{spectral}
one can see  that there are good agreements between our analytical results and the results obtained from the full numerical analysis at the classical level. Finally, $n_f \simeq 7 $
which is in agreement with the numerical results (see {\bf Fig. }\ref{stf}, $N_c\simeq 56$ and $N_e \simeq 63.5$ so $n_f \simeq 7$).

\subsection{Quantum Fluctuations Back-reactions}
\label{back-reaction}

Our analysis so far concentrated on the classical evolution of the system. Due to tachyonic instability during the phase transition quanta of $\psi$ particles inhomogeneities, $\delta \psi ({\bf x} , t)$,  are produce  
\cite{Felder:2000hj, Felder:2001kt}
which  can back-react on the classical backgrounds as in preheating models 
\cite{Kofman:1994rk, Traschen:1990sw, Shtanov:1994ce}.
If the mechanism of particle creation due to tachyonic instability is very efficient, the back-reaction of the produced particles can induce large mass for inflaton field in the form of $g^2 \langle \delta \psi^2 \rangle$ where $\langle \delta \psi^2 \rangle$ is the expectation value of 
$\delta \psi$ in the Hartree approximation. This violates the slow-roll conditions and $\phi$ rapidly rolls towards the global minimum. Furthermore, the large back-reactions of $\delta \psi$
also induce a large mass for the background $\psi$  which can uplift its tachyonic mass. 
Therefore, one has to take into account the quantum back-reactions and see which of the classical or quantum back-reactions dominate first to terminate inflation. 

To handle the quantum back-reactions, we use the Hartree approximation and calculate the 
effects of all modes which become tachyonic during the phase transition. 
Let us compute which modes become tachyonic after the phase transition. The equation of the 
$\delta \psi_k$ fluctuations at the linear order is 
\ba
\delta \psi_k'' + 3 \delta \psi_k' + \left(\frac{k^2}{a^2 H^2}  - 2 \beta r n \right) \delta \psi_k =0
\ea
Therefore modes which satisfy the inequality 
\ba
\dfrac{k}{k_c}  \lesssim \sqrt{2\beta \,r} = \epsilon_{\psi} 
\ea
become tachyonic soon after the phase transition. Here we defined $k_c \equiv a(N_c) H_0 = e^{N_c} H_0$ as the critical mode which leaves the horizon at the time of waterfall phase transition.

We divide the tachyonic modes into two categories: first,  large modes $\psi_k^L$, corresponding to $k<k_c$, which exit the Hubble radius sometime before the phase transition and second, the small modes $\psi_k^S$, corresponding to $k_c < k < \epsilon_{\psi} k_c$, 
which do not exit the Hubble radius till time  of phase transition. 
Using { Eq.} (\ref{qem}), the amplitude of both modes at the time of critical point $\phi = \phi_c$ is
\ba
|v_k(n=0)| \simeq  \dfrac{1}{\sqrt[4]{|4k^2-2/\eta^2|}} \, ,
\ea
and therefore for large modes one has 
\ba
\label{delpsi-ls}
|\delta \psi _k^L (n=0)| \simeq  \dfrac{e^{-3N_c/2}}{\sqrt{2H}} \, .
\ea
This is in agreement with our previous result, {Eq.}(\ref{delsf}), for $n=0$. On the other hand, for the small modes which remain sub-horizon till the time of phase transition  one has
\ba
\label{delpsi-ss}
|\delta \psi _k^S (n=0)| \simeq  \dfrac{e^{-N_c}}{\sqrt{2k}} \, .
\ea

With this division of the modes, and using { Eq.} (\ref{delsf}), 
$\langle \delta \psi^2 \rangle$  is calculated to be
\ba
\langle \delta \psi ^2 \rangle & \simeq& \left[  \int_{0}^{k_c} \dfrac{\mathrm{d}^3 k}{(2\pi)^3} \dfrac{1}{2H_0}\,e^{- 3 N_c}
+ \int_{k_c}^{\epsilon_{\psi} k_c} \dfrac{\mathrm{d}^3 k}{(2\pi)^3} \dfrac{1}{2k}\,e^{- 2 N_c}
 \right]~ \exp \left( -3 n +\dfrac{4\epsilon_{\psi}}{3} n^{3/2} \right) 
\nonumber\\
&\equiv& \langle \delta \psi ^2 \rangle_L + \langle \delta \psi ^2 \rangle_S \, ,
 \label{quantum-psi}
\ea 
in which we have ignored factors of order unity. The first integral, representing 
$\langle \delta \psi ^2 \rangle_L$,  comes from the large modes which leave the horizon sometimes prior to phase transition whereas the second integral,  representing $\langle \delta \psi ^2 \rangle_S$, indicates the contributions from the very small scales modes which are sub-horizon till phase transition.

We observe that both $\langle \delta \psi ^2 \rangle_S $ and 
$\langle \delta \psi ^2 \rangle_L $ have similar $n$-dependence which becomes important in our discussion below when we compare the classical and quantum back-reactions. However, 
\ba
\label{StoL}
\langle \delta \psi ^2 \rangle_S  \sim \epsilon_{\psi}^2   \langle \delta \psi ^2 \rangle_L \, . 
\ea
As $\epsilon_{\psi}^2 \gg 1$ the second integral in Eq. (\ref{quantum-psi}) is much bigger than the first one. 
This means that the cumulative contributions of the  modes which becomes tachyonic but remained sub-horizon till the time of phase transition are more important in quantum back-reactions and  
\ba
\label{q-back}
\dfrac{\langle \delta \psi ^2 \rangle}{H_0^2} \simeq 
\dfrac{\langle \delta \psi ^2 \rangle_S}{H_0^2}
\sim \dfrac{\epsilon^2_{\psi}}{4 \pi ^2}  ~ \exp \left( \dfrac{4\epsilon_{\psi}}{3} n^{3/2} -3n \right)  \, .
\ea

As explained before, there are two competitive corrections which terminate inflation and the evolution of large scale curvature perturbations during the phase transition. The First one is  the  classical corrections from the  $g^2 \phi^2\psi^2/2$  interactions which induce large mass for the inflaton, violating the slow-roll condition during the phase transition. This is the effect which was studied in subsection (\ref{classical}), specifically Eq. (\ref{c-back}).
The second contribution is the quantum back-reaction corrections into inflaton mass via
$g^2 \phi^2 \langle \delta \psi ^2 \rangle$. We need to see which of the above two corrections dominate first.  Comparing { Eq.} (\ref{c-back}) and { Eq.} (\ref{q-back}), one finds that these two terms have similar time-dependence ($n$-dependence) and 
\ba
\label{qtoc}
\dfrac{\langle \delta \psi^2 (n)\rangle}{\psi^2(n)} 
\simeq \dfrac{\langle \delta \psi^2 (n)\rangle_S}{\psi^2(n)}
\simeq \epsilon_{\psi}^2 \,\dfrac{g^2}{\beta} \left(\dfrac{\phi_i}{\psi_i} \right)^2 \,e^{3N_c}\, .
\ea
Using { Eq.}(\ref{vd-ne}) and { Eq.}(\ref{psi-phi}), one finds that
\ba
\label{q-c}
\dfrac{\langle \delta \psi^2(n) \rangle}{\psi^2(n)} \gg  \left( g^2 \beta^{3/2} \right) e^{3N_c} \, .
\ea
As $\exp(3N_c)\sim 10^{78}$, one concludes that  for any reasonable value of coupling $g$  
the quantum back-reactions dominate completely over the  classical back-reactions to terminate inflation. Therefore,  the end-point of inflation is determined by the quantum back-reaction effects from 
$\langle \delta \psi^2 (n)\rangle_S$ and 
\ba
\label{eoi}
\exp \left( \frac{4}{3} \, \epsilon_{\psi} \,n_f^{3/2} -3 n_f  \right) \simeq \, \dfrac{8 \pi^2 \alpha}{g^2 \epsilon_{\psi}^2}.
\ea

 It's now instructive to calculate $n_f$, the time when the quantum back-reactions terminate inflation. If we follow the estimation of \cite{Felder:2000hj, Felder:2001kt} performed in flat backgrounds, we obtain 
\ba
n_f  \sim \dfrac{1}{\epsilon_{\psi}} \ln \dfrac{8 \pi^2 \alpha}{g^2} \, .
\ea
For the  parameters of our numerical investigations such as in {\bf Fig  \ref{bkd}} one has 
$n_f  \simeq 3$. However,  taking into account the background cosmological expansion, from our Eq. (\ref{eoi}), one finds
\ba
\label{nf-final}
n_f \sim \left( \dfrac{1}{\epsilon_{\psi}} \ln \dfrac{8\pi^2\,\alpha}{\epsilon^2_{\psi}\,g^2} \right)^{2/3} \, ,
\ea
which for the parameters of our numerical studies this gives $n_f \simeq 1.7$. This clearly
indicates that quantum back-reactions dominate sooner than the classical back-reaction which
happens at $n_f \simeq 7$.
 
Another important point which should be considered is that the quantum fluctuations, 
$\langle \delta \psi^2 \rangle$, can change the effective classical trajectory. 
As we demonstrated in Eq. (\ref{qtoc}) the 
expectation value of the quantum fluctuations dominates completely over  the ``zero momentum" mode  and they will induce new effective trajectory in the field space. 
The fact that $\langle \delta \psi^2 \rangle$ dominates over the background classical field contributions in waterfall dynamics also indicates that  using the $\psi $  ``zero momentum'' as the classical trajectory is not reliable. Quantum modes with tachyonic mass become highly occupied soon after the phase transition and in some senses they become classical.
This suggests that we may take $\sqrt{ \langle \delta \psi^2 \rangle}$ as the effective classical field trajectory. Below we justify this proposal so  one can introduce an effective classical trajectory defined by $\psi \rightarrow \sqrt{ \langle \delta \psi^2 \rangle} $ . 

To justify our proposal, we start with the conventional method for the evolution of  super-horizon curvature perturbations \cite{Wands:2000dp}. In the study of curvature perturbations
in hybrid inflation this method was pioneered in  \cite{Lyth:2010ch} and 
\cite{Lyth:2010zq} (see also \cite{Gong:2010zf}). In this method, the change in the comoving curvature perturbation on super-horizon scales is given by 
\ba
\label{R-dot-eq}
\dot{{\cal R}}_{c\K}  =  H  \dfrac{\delta p_{c\K}}{\rho + p} \, ,
\ea
where $\delta p_{c}$ is the pressure perturbations  on comoving slices. In our formalism 
we relate $\delta p_{c}$ to the entropy perturbations $\delta s$ or $\delta \psi$.
In the analysis here it is assumed that the $\psi$ field is frozen at the background so $\psi=0$
classically.  Our goal is to demonstrate that on super-horizon scales Eq. (\ref{R-dot-eq}) 
reduces to our starting equation
for the evolution of curvature perturbation,  Eq. (\ref{dotR}), obtained in \cite{Gordon00} 
for the two field inflationary system, with the appropriate definition of $\dot \theta$ and replacing $ \psi \rightarrow \sqrt{\langle \delta \psi^2 \rangle}$.

One can simply check  that $\rho +p = \dot{\sigma}^2$, where taking into account the quantum effects, one also has \cite{Gong:2010zf}
$\dot{\sigma}^2 = \dot{\phi}^2 + \langle \dot{\delta \psi}^2 \rangle$.  However, as in our previous analysis,  the velocity along the inflation trajectory in the field space 
can be well approximated by $\dot{\sigma}^2 = \dot{\phi}^2$ till end of inflation. Therefore 
\ba
\label{Rdot-sim}
\dot{{\cal R}}_{c\K}  =  \dfrac{ H }{\dot{\sigma}} \dfrac{\delta p_{c\K}}{\dot{\sigma}}.
\ea
 As the potential is vacuum dominated we can also neglect the gravitational back-reactions on $\delta \psi$. Neglecting the self interaction term,  and noting that the comoving slice  coincides with  the $\delta \phi =0 $ surface \cite{Gong:2010zf},  the contribution of  $\psi$-field in the pressure becomes
\ba
\delta p_{c } = \dfrac{1}{2} \dot{\delta \psi}^2 + \dfrac{1}{2} (M^2 - g^2 \phi^2) \delta \psi^2 \, .
\ea
Using  { Eq.} \ref{delpsi-n} for the time evolution of $\delta s$ mode (which is the same as 
 $\delta \psi$ mode)  one finds that the contribution of the kinetic term above is approximately equal to the potential term and 
 \ba
 \label{delta-p}
\delta p_{c\K} \simeq (M^2 - g^2 \phi^2) \left( \delta \psi^2 \right)_{\K} \, .
\ea 

Below we would like to find an expression for $\left( \delta \psi^2 \right)_{\K}$ to calculate $\delta p_{c\K}$. One can show that \cite{Gong:2010zf}
\ba
\langle \left(\delta \psi^2 \right)_{\K} \left(\delta \psi^2 \right)_{\K'}\rangle = 2 \int \md ^3 p \vert \delta \psi^2 _{p} \vert \vert \delta \psi^2 _{\vert \K- \mathbf{p} \vert} \vert \delta^3(\K+\K') \, .
\ea
Most of the contributions to the momentum $p$ integration above comes from the small scales modes with the amplitude 
\ba
\delta \psi_{\K} (n) = \delta \psi_{\K} (n=0) f(n)\, ,
\ea
in which $f(n)$ is defined via 
\ba
f(n) \equiv \exp \left( \dfrac{2 \epsilon_{\psi}}{3} n^{3/2} -3 n \right) \, ,
\ea
and $\delta \psi_{\K} (n=0) = e^{-N_c}/\sqrt{2 k}$ as given by Eq. (\ref{delpsi-ss}). Performing the integral and setting the UV cutoff $p= \epsilon_{\psi} k_c$ yields 
\ba
\label{A1}
\langle \left(\delta \psi^2 \right)_{\K} \left(\delta \psi^2 \right)_{\K'}\rangle &=& (2 \pi)^3 \delta^3(\K+\K')  \int \md p \,p^2  \vert \delta \psi^2 _{p} \vert \int \dfrac{e^{-2N_c}}{2 \pi^2} \dfrac{- \md \cos \theta}{(k^2+p^2-2kp \cos \theta)^{1/2}} ~f^4(n) \nonumber
\\
&=& (2 \pi)^3 \delta^3(\K+\K') \dfrac{e^{-2N_c}}{2 \pi^2} \int  \md p\, p  \vert \delta \psi^2 _{p} \vert~ f^4(n) \nonumber\\
&=& (2 \pi)^3 \delta^3(\K+\K') \dfrac{e^{-3N_c}}{4 \pi^2} \epsilon_{\psi} H_0 ~f^4(n) \, .
\ea
On the other hand, for the  large scale  quantum fluctuations we have 
\ba
\label{A2}
\langle \delta \psi_{\K}^L \delta \psi_{\K'}^L\rangle = (2 \pi)^3 \delta^3(\K+\K') \dfrac{e^{-3N_c}}{2H_0 } f(n)^2 \, ,
\ea
where Eq. (\ref{delpsi-ls}) have been use for their amplitudes at the time of waterfall. Furthermore, as we showed in Eq. \ref{q-back} 
\ba
\label{A3}
\langle \delta \psi^2 \rangle = \dfrac{H_0^2}{4\pi^2} \epsilon_{\psi}^2 f(n)^2 \, .
\ea
Combining Eqs. (\ref{A1}), (\ref{A2}) and (\ref{A3}) we obtain
\ba
\langle \left(\delta \psi^2 \right)_{\K} \left(\delta \psi^2 \right)_{\K'}\rangle =
 \frac{2}{\epsilon_\psi} \langle \delta \psi^2 \rangle  \langle \delta \psi_{\K}^L \delta \psi_{\K'}^L\rangle  \, .
\ea
This equation suggests that, as long as we are interested in two-point functions at the linear perturbation theory, one can make the following identification
\ba
\left(\delta \psi^2 \right)_{\K} \rightarrow \sqrt{\dfrac{2}{ \epsilon_{\psi}}} 
 \sqrt{\langle \delta \psi^2 \rangle} \delta \psi_{\K}^L  \, .
\ea
Having obtained this identification for $\left(\delta \psi^2 \right)_{\K}$ we plug it into 
$\delta p_{c\K}$ expression in Eq.  (\ref{delta-p}) and Eq. (\ref{Rdot-sim}) to obtain
\ba
\label{Rdot-subs}
\dot{{\cal R}}_{c\K}  =  \dfrac{ 2 H }{\dot{\sigma}} \sqrt{\dfrac{1}{2 \epsilon_{\psi}}}
\dfrac{(M^2 -g^2 \phi^2) \sqrt{\langle \delta \psi^2 \rangle}}{\dot{\sigma}} \delta \psi^{L}_{\K}. 
\ea
 Now we can cast  Eq. (\ref{Rdot-subs}) in the form of our starting formula  Eq. (\ref{dotR})
for $\dot{{\cal R}}$ as prescribed in\cite{Gordon00}. Noting that $(M^2 -g^2 \phi^2) \sqrt{\langle \delta \psi^2 \rangle} = -V_{,\psi}(\psi \rightarrow \sqrt{\langle \delta \psi^2 \rangle})$ and $\delta s_{\K} = \delta \psi_{\K}$ on super-horizon scales,  Eq. (\ref{Rdot-subs}) can be written as 
\ba
\label{Rdot-subs2}
\dot{{\cal R}}_{\K}  =  \dfrac{ 2 H }{\dot{\sigma}} \dot{\theta}_{\mathrm{eff}}  \delta s_{\K} ~ \sqrt{\dfrac{1}{2 \epsilon_{\psi}}}
\ea
where similar to Eq. (\ref{dot-theta-eq})
\ba
\dot{\theta}_{\mathrm{eff}} = - \dfrac{V_s \left(\psi \rightarrow \sqrt{\langle \delta \psi^2 \rangle} \right)}{\dot{\sigma}} \, .
\ea
Therefore we have justified our proposal in using   $\sqrt{\langle \delta \psi^2 \rangle}$ as the effective trajectory when the background  $\psi$ field is zero. Compared to our starting formula 
Eq. (\ref{dotR}), there is an extra factor $\sqrt{1/2\epsilon_{\psi}}$ in Eq. (\ref{Rdot-subs2}) which originated from the UV cutoff imposed in Eq. (\ref{A1}). However, this extra factor is
not significant and it does not affect our analysis below. For instance, in our numerical example we have  $\sqrt{1/2\epsilon_{\psi}} \simeq 1/4$.

The quantum back-reactions therefore have two crucial effects. First they determine the end of inflation given by Eq.  (\ref{qtoc}) and { Eq.} (\ref{eoi}). Second, as demonstrated above, they provide the effective classical trajectory via  
$\psi \rightarrow \sqrt{ \langle \delta \psi^2 \rangle} $.
To see how the replacement $\psi \rightarrow \sqrt{ \langle \delta \psi^2 \rangle} $  changes our results in previous sections based on classical analysis we note that the only place in which we need the classical trajectory is in the calculation of $\theta '$ in the { Eq.} (\ref{R-eq}). However, $\psi$ and $\sqrt{ \langle \delta \psi^2 \rangle }$ have the same time dependence and they only differ by a 
normalization which is given by { Eq.} (\ref{qtoc}). One can simply adopt our previous results, Eq. (\ref{Rnf2}), for the final curvature perturbations but with $\sqrt{ \langle \delta \psi^2 \rangle }$ as the classical trajectory and take into account the extra modifying factor $\sqrt{1/2\epsilon_{\psi}}$ mentioned above.
Using { Eq.} (\ref{qtoc}) and { Eq.} (\ref{eoi}) in Eq. (\ref{Rnf2}), the amplitude of curvature perturbation therefore is

\ba
\label{Rnf4}
{\cal R}_f \simeq{\cal R}_0\left [ 1- \left( \dfrac{k}{k_c}\right)_{\ast}^{3/2} \left(\dfrac{24 \pi \epsilon_{\psi}^{-5/6} }{ g \beta^{3/4}} \right)  \right] \, ,
\ea
in which we considered $\tilde{\beta} \simeq \beta$, $k_c = e^{N_c} H_0$ and neglected the $r$ dependence in the exponent.
Similar to the analysis in classical case, the induced curvature perturbations becomes 
\ba
\label{final}
\frac{\Delta {\cal P_ R} }{{\cal P_R}_0}\sim \dfrac{\epsilon_{\psi}^{-5/3}} {g^2 \beta^{3/2} }  \left( \dfrac{k}{k_c}\right)^3\, .
\ea
As in \cite{Lyth:2010ch,Fonseca:2010nk}, Eq. (\ref{final}) shows that the induced curvature perturbations from the entropy perturbations 
has the power spectrum $\propto k^3$ and is suppressed compared to the primordial curvature perturbations by the factor $e^{-3N_c}$.
As $N_{\ast} \ll N_c$ and $e^{- 3 N_c} \sim 10^{-78}$ one concludes that $\frac{\Delta {\cal P_ R} }{{\cal P_R}_0} \ll 1$ for any reasonable values of the coupling $g$. This indicates that there would be no large scale curvature perturbations once the quantum back-reactions are turned on during the phase transition. 
Physically this means that the cumulative quantum back-reactions of very small scale modes, modes which become tachyonic during the phase transition but remained sub-horizon during entire inflation, shuts off the background $\psi$  instability, forcing $\phi$ and $\psi$ to their global minima ending inflation quickly. The tachyonic instability of entropy perturbations are lifted and they can not produce appreciable large scale curvature perturbations during the phase transition. Our results can be compared with the findings of \cite{Levasseur:2010rk, Barnaby:2006km, Mazumdar:2010sa}, see also \cite{Enqvist:2004ey, Enqvist:2004bk}, although these works concern about the second order curvature perturbations effects. 

We can also compare our results for $\Delta {\cal P_R}$ with the result obtained in \cite{Gong:2010zf} using the  $\delta N$ formalism.
Noting that ${\cal P_R}_0 \simeq  g^2 / 4 \pi^2 \beta r^2$, we obtain
$\Delta {\cal P_R} \sim \epsilon_{\psi}^{-20/3} ( \dfrac{k}{k_c})^3 $
whereas in \cite{Gong:2010zf} they have $\Delta {\cal P_R} \sim \epsilon_{\psi}^{-22/3} (\frac{k}{k_c})^3$. The agreements between these two results are good (the two estimations of $\frac{\Delta {\cal P_ R} }{{\cal P_R}_0}$ differ by one order of magnitude in $ 3 N_c \sim 78$ orders of magnitude). 

One concern may be the overproduction of primordial black holes in this model. This question was studied in \cite{GarciaBellido:1996qt}. As they showed, with $\beta \gg 1$, the standard hybrid inflation model is safe under overproduction of primordial black holes.

\section{Conclusions}
\label{conclusions}

In this paper the possibility of producing large scale curvature perturbations induced from the entropy perturbations during the waterfall phase transition in hybrid inflation are studied. We have shown that 
whether or not appreciable amounts of large scale curvature perturbations  are produced depend crucially on the competition between classical and quantum mechanical back-reactions to terminate inflation.  If one considers only the classical back-reaction effects, one obtains a significant large scale curvature perturbations which completely dominate over the initial curvature perturbations. The induced large scale curvature perturbations would be highly blue-tilted with $n_{\cal R} \simeq 4$ as in \cite{Gong:2008ni}. However, we have shown that the quantum-mechanical back-reactions of the waterfall field inhomogeneities produced during the phase transition dominate before the classical-back-reaction becomes important. In the Hartree approximation we found that the quantum back-reactions shuts off the classical tachyonic instability very efficiently  terminating inflation as well as curvature perturbations evolutions quickly after phase transition. We have shown that the main contribution to quantum back-reactions comes from the cumulation of the very small scales inhomogeneities, modes which are tachyonic during the phase transition but remain sub-horizon during entire inflationary period. We also made the interesting observation that in standard hybrid inflation where 
the waterfall field rapidly freezes to $\psi=0$ at the background level, one can use $\psi \rightarrow  \sqrt{ \langle \delta \psi^2 \rangle} $ as the effective classical trajectory. 
In summary, the quantum back-reactions have two crucial effects. First they determine the end of inflation given by Eq. (\ref{qtoc}) and { Eq.} (\ref{eoi}). Second, as mentioned above, they provide the effective classical trajectory via  $\psi \rightarrow \sqrt{ \langle \delta \psi^2 \rangle} $ .

Although we have presented the analysis here only for the standard hybrid inflation, but we believe that this picture also holds for other models of inflation where there are sharp phase transitions at the end of inflation. This includes models of  brane inflation where inflation ends abruptly due to 
tachyon formation once the distance between the brane and anti-brane reaches a critical value.
However, it would be interesting to see what happens in models of inflation, such as in double inflation \cite{Silk:1986vc} where there is a mild phase transition in fields evolution during early stages of inflation. Since the difference $N_c - N_*$ in Eq. (\ref{final}) is not very large, an appreciable amount of curvature perturbations can be created even when the quantum mechanical back-reactions are taken into account. This in turn can produce features in curvature perturbations such as in models  \cite{Joy:2007na, Battefeld:2008py,  Battefeld:2010rf}.  It would be interesting to see the observational effects of the phase transitions during inflation as considered e.g. in \cite{Parkinson:2004yx, Joy:2008qd}.

 \vspace{1 cm}

{\bf{Acknowledgments}}

\vspace{1 cm}
We would like to thank Razieh Emami, Jinn-Ouk Gong, David Lyth, Mohammad Hossein Namjoo, David Wands  and specially M. Sasaki  for useful discussions and correspondences. 
We specially thank Bruce Bassett for many insightful discussions and for bringing Refs. \cite{Tsujikawa:2002nf, Tsujikawa:2002qx, Tanaka:2003cka} into our attention which initiated this work. A. A. A. would like to thank  IPM and ``Bonyad Nokhbegan Iran'' for partial support.


\appendix
\section{Properties of Airy functions}
Here we summarize some important properties of the Airy functions which are used in the main text.

To the second leading order the Airy functions of second kind and their derivatives have the following asymptotic expansion for large arguments
\ba
\mathrm{Bi}(z) \sim \pi^{-1/2} z^{-1/4} e^{\zeta} \left( 1+ c_1 \zeta ^{-1} \right)\\
\mathrm{Bi}'(z) \sim \pi^{-1/2} z^{1/4} e^{\zeta} \left( 1+ d_1 \zeta ^{-1} \right)\\
\ea
in which 
\ba
\zeta = \dfrac{2}{3} z ^{3/2}
\ea	
and $c_1= 5/72$ and $ d_1= -7/5 ~c_1$.
By using the above equations one can estimate the derivative of Airy functions to the second leading order as
\ba
\mathrm{Bi}'(z) \simeq \sqrt{z}~ \mathrm{Bi}(z) \left( 1- \dfrac{1}{4 z^{3/2}}\right) \, .
\ea
By using the above approximation for the derivative of Airy function one can find 
\ba
\int \mathrm{Bi}(a\,z)^2 \md z \simeq \dfrac{1}{2\,a^{3/2}} (z)^{-1/2} ~ \mathrm{Bi}(a\,z)^2 \, ,
\ea
and  also
\ba
\int  z~\mathrm{Bi}(a\,z)^2 \md z \simeq \dfrac{1}{2\,a^{3/2}} (z)^{1/2} ~ \mathrm{Bi}(a\,z)^2 \, .
\ea
Similarly, one can show that in the leading order the following relation also holds
\ba
\int z^n \mathrm{Bi}(a\,z)^2 \md z \simeq \dfrac{1}{2\,a^{3/2}} (z)^{n-1/2} ~ \mathrm{Bi}(a\,z)^2 \, .
\ea
Also one can find the following useful relation
\ba
\label{mixint}
\int z^n~e^{\alpha \, z} \mathrm{Bi}(a\,z)^2 \md z \simeq \dfrac{e^{\alpha \,z}}{2\,a^{3/2}} (z)^{n-1/2} ~ \mathrm{Bi}(a\,z)^2 \, .
\ea
which was used to estimate the integral in Eq. (\ref{In1}).

Finally there is another simple but important relation for second derivative of Airy functions
\ba
\label{sdaf}
\mathrm{Bi}''(z)=z \mathrm{Bi}(z) \, .
\ea
  
\vspace{0.5cm}


\end{document}